\documentclass[12pt]{iopart}
\usepackage{latexsym}
\usepackage{graphicx}
\usepackage{amssymb}
\usepackage[english]{babel}
\newcommand{\bq}{\begin{equation}}
\newcommand{\eq}{\end{equation}}
\newcommand{\ba}{\begin{eqnarray}}
\newcommand{\ea}{\end{eqnarray}}

\begin{document}
\title{Pair-factorized steady states on arbitrary graphs}
\author{B. Waclaw$^1$, J. Sopik$^2$, W. Janke$^1$ and H. Meyer-Ortmanns$^2$}
\address{$^1$Institut f\"ur Theoretische Physik,Universit\"at Leipzig, Postfach 100\,920, 04009 Leipzig, Germany}
\address{$^2$School of Engineering and Science, Jacobs University, P. O. Box 750561, 28725 Bremen, Germany}
\ead{h.ortmanns@jacobs-university.de}

\begin{abstract}
Stochastic mass transport models are usually described by specifying hopping rates of particles between sites of a given lattice, and the goal is to predict the existence and properties of the steady state. Here we ask the reverse question: given a stationary state that factorizes over links (pairs of sites) of an arbitrary connected graph, what are possible hopping rates that converge to this state?
We define a class of hopping functions which lead to the same steady state and guarantee current conservation but may differ by the induced current strength.
For the special case of anisotropic hopping in two dimensions we discuss some aspects of the phase structure.
We also show how this case can be traced back to an effective zero-range process in one dimension which is solvable for a large class of hopping functions.
\end{abstract}
\pacs{89.75.Fb, 05.40.-a, 64.60.Ak}
\maketitle

Stochastic transport of some conserved quantity, generically called
``mass'', has recently attracted much attention due to a large variety of
applications ranging from microscopic (intracellular)  to
macroscopic (highway) traffic \cite{chowdhury}.
Important examples of technological interest are granular flow \cite{liu}
and granular clustering \cite{majumdar}. From a theoretical point of view,
these systems are challenging, since they are in general
out-of-equilibrium and allow for phase transitions even in one
dimension \cite{evans2000}.
An example is spontaneous symmetry breaking and the phenomenon of condensation which happens above some critical
mass density and corresponds to jams in traffic or
aggregation in granular media.
Nevertheless, simple models such as the zero-range process (ZRP) \cite{2006}, a dynamical version of the balls-in-boxes model \cite{bbj}, or asymmetric simple exclusion processes \cite{asep}, can capture some important aspects of these problems while remaining analytically solvable.

The non-equilibrium models are defined by specifying the dynamics rather than the probability of a microstate. Usually one proposes the transition rates between states, and the goal is to predict the existence and the properties of a stationary state. This is the state where macroscopic observables remain constant, although some currents may flow in the system. In this paper we study the reverse problem. Given a steady state which assumes a form factorized over pairs of sites that correspond to the links of an arbitrary graph, we search for a class of transition probabilities which lead to this state. This approach is motivated by the fact that knowledge of the stationary states of non-equilibrium models generically facilitates the discussion of the phase diagram, because a number of observables can be calculated analytically.
For instance, the ZRP, defined in terms of particles hopping between sites of a lattice and interacting only if they
are at the same node, has a steady state that factorizes over
the sites of a lattice, or more generally, over nodes of an arbitrary graph.
The factorization allows for a convenient mathematical treatment.
A generalization of the ZRP that leads to pair-factorized steady states (PFSS) was proposed in \cite{evans1} for a one-dimensional ring topology.
It was shown that nearest-neighbor exponentially suppressed interactions plus some extra ``pinning'' (ZRP-like) potential result in a condensate that is spatially extended.
In \cite{bw1} the shape of the condensate was derived. It was also shown that the scaling of the extension of the condensate with the system size can be tuned via an appropriate competition between local and ultralocal hopping interactions.

We shall show in this paper that the ZRP and the PFSS on a ring (where PFSS here should be understood as the corresponding processes leading to PFSS)
are special cases of a more general setting.
Beyond that, we shall consider non-local processes on an arbitrary graph, or processes with anisotropic hopping in two dimensions. In the latter case we shall show that it can be dimensionally reduced to a ZRP in one dimension with weights that contain the information on the pair-factorized stationary behavior in the second dimension. Therefore former results on PFSS on a one-dimensional ring topology \cite{bw1} can be used to derive features of the condensation transition in the anisotropic case.

{\bf The model.}
We consider a connected, undirected but otherwise arbitrary graph with $N$ nodes (sites), and node degrees $k_1,...,k_N$. We place $M$ particles of unit mass on the nodes of this graph. ``Particles'' here stands for a generic mass that is involved in the transport process and has ``bosonic'' properties in the sense that $m_i\geq 0$ particles may be assigned to the same site $i$. The distribution of occupation numbers of nodes is denoted as $\vec{m}=\{m_1,...,m_N\}$.
The dynamics is defined as follows.
We pick up a randomly chosen node $i$, and if it is not empty, a single particle departures with probability $u_i(m_i|\dots)$, where the dots stand for occupation numbers of other sites (in general not necessarily nearest neighbors of node $i$).
Next, the particle chooses a target site $j$ with probability $W_{ij}\equiv W(i\to j)$. The transition matrix $W$ may be arbitrary, with the only assumption that all $W_{ij}\geq 0$ and $\sum_j W_{ij}=1$ for any $i$.
The hopping event is thus split into two steps: the departure from a site, determined by the function $u$, and the choice of destination site, determined by the rates $W_{ij}$.

We will next derive under which conditions on the hopping rate and the transition matrix the system reaches a steady state that assumes a pair-factorized form
\bq
	w(\vec{m})=\prod_{\left<i,j\right>}g_{ij}(m_i,m_j) \prod_{i=1}^{N}\;k_i^{m_i}\delta\left(\sum_{i=1}^{N}m_i-M\right).
	\label{eq1}
\eq
Here $\left<i,j\right>$ denotes pairs of nearest neighbors, $g_{ij}(m_i,m_j)$ is a symmetric but otherwise arbitrary weight function that may depend on the link $(i,j)$, and the $\delta$-function  that ensures conservation of the overall mass $M$ and will be dropped below.
The factors $k_i^{m_i}$ are slit off for convenience.
As we shall prove next, a possible class of hopping rates (although not the only possible one, since, for example, the symmetry condition on $g(m,n)$ was released in \cite{evans1}) is given by
\bq
	u_i(m_i|\dots) = \prod_{j\in \mathcal{N}(i)}\frac{g_{ij}(m_i-1,m_j)}{g_{ij}(m_i,m_j)},
	\label{eq4}
\eq
where $\mathcal{N}(i)$ denotes the set of neighbors of node $i$
provided the transition matrix $W_{ij}$ satisfies the condition
\begin{equation}
	k_i = \sum_{j} W_{ji} k_j\;. \label{kwij}
\end{equation}
{\bf Proof. } Let us start with the balance equation for a conserved probability current at each node $i$, this is a condition that is necessary for having a steady state:
\ba
	u_i(m_i|\dots) w(\vec{m}) &=&\sum_{j}W(j\to i) u_j(m_j+1|\dots,m_i-1,\dots) \nonumber \\
	&\times & w(m_1,\dots,m_i-1,\dots,m_j+1,\dots,m_N)\;.
	\label{5bw2}
\ea
Equation (\ref{5bw2}) says that the probability with which a particle leaves a site $i$ with $m_i$ particles in a configuration given by occupation numbers $\vec{m}$ should be the same as the total probability that node $i$ receives one particle from any site $j$ that was formerly in a configuration with $m_j+1$ particles at site $j$ and $m_i-1$ particles at site $i$. The individual probabilities for a hopping event from $j$ to $i$ are given by $W(j\rightarrow i)$. As a sufficient condition, the probability current is conserved if (\ref{5bw2}) is satisfied individually for any $i\in \{1,..,N\}$ and any set of occupation numbers  $\vec{m}$.
Now, dividing (\ref{5bw2}) by $w(\vec{m})$, using the definition (\ref{eq1}) of the steady state, and inserting (\ref{eq4}) as an assumed form of $u_i(m_i|\dots)$, we cancel common factors in the ratio $w(...)/w(...)$ on the right-hand side and expand it to obtain 
\ba
	u_i(m_i|\dots) = \sum_{j} W_{ji} \left(\frac{g_{ij}(m_j,m_i-1)}{g_{ij}(m_j+1,m_i-1)}  \prod_{\left.\right.^{a\in \mathcal{N}(j)}_{a\neq i}} \frac{g_{aj}(m_j,m_a)}{g_{aj}(m_j+1,m_a)}\right) \frac{k_j}{k_i} \nonumber \\
	\times \left(\prod_{\left.\right.^{b\in \mathcal{N}(i)}_{b\neq j}} \frac{g_{bi}(m_i-1,m_b)}{g_{bi}(m_i,m_b)}\right) \frac{g_{ij}(m_i-1,m_j+1)}{g_{ij}(m_i,m_j)}
	\prod_{\left.\right.^{c\in \mathcal{N}(j)}_{c\neq i}} \frac{g_{cj}(m_j+1,m_c)}{g_{cj}(m_j,m_c)} .
\ea
Note that the first bracket is just $u_j$, rewritten as a product of ratios of $g$, with the part for $a=i$ written out explicitly. Similarly we express the ratio of $w(...)/w(...)$ in terms of the weight functions $g$ and explicitly split off the factor for $b=j$. The first product (over $a$) and the last product (over $c$) as well as the terms $g_{ij}$ cancel out. We end up with
\bq\label{eq8}
	u_i(m_i|\dots) = \sum_{j} W_{ji} \frac{k_j}{k_i} \left(\prod_{b\in \mathcal{N}(i)} \frac{g_{bi}(m_i-1,m_b)}{g_{bi}(m_i,m_b)}\right)\;,
\eq
where the product over $b$ now includes all nearest neighbors of $i$. The expression in the bracket is just $u_i(m_i|\dots)$. Therefore, equation (\ref{eq8}) is satisfied if
\bq\label{eq9}
	k_i = \sum_{j} W_{ji} k_j \;.
\eq
This is precisely (\ref{kwij}) which completes the proof. $\square$

Before we come to examples, we want to add a remark on a useful possible reinterpretation of the involved links included in $W_{ij}$.
Since the stationary state (\ref{eq1}) only refers to the occupation number distribution and depends on the hopping function $u$, but not on the transition matrix $W_{ij}$, by making different choices on $W_{ij}$ we can tune the current of particles in the system. Moreover, the set of non-zero $W_{ij}$ does not need to coincide with the set of existing links in the graph. This may be seen as if we had two graphs: an undirected one that specifies which pairs of nodes enter the steady state (\ref{eq1}), and a directed and weighted graph of allowed transitions, defined by $W_{ij}=W(i\to j)$.
This fact can be used in numerical simulations of the steady state. Suppose that we have a process which, although being out of equilibrium, leads to (\ref{eq1}). If we are interested only in static properties of the steady state and have to turn to Monte Carlo simulations as the model is not fully analytically accessible, we can change the dynamics (by changing $W_{ij}$) in order to speed up the convergence towards the stationary state, or to reduce the autocorrelation time. An example is to turn the local hopping of particles with given hopping rates into a non-local update scheme by choosing two random nodes and moving the particle between them, accepting or rejecting the move according to the rules of the Metropolis algorithm. This can be viewed as specifying $W_{ij}=1/(N-1)$ for all $i,j$, i.e, as making the ``transition'' graph a complete graph. One can check that such a choice obeys (\ref{kwij}), and since the probability of moving the particle along each link is the same in both directions, there is no macroscopic particle current. More generally, if $W_{ij}=1/k_i$ for every link $(i,j)$, equation (\ref{kwij}) is fulfilled and the current vanishes, so the system is actually in equilibrium.
This method can in particular be used to simulate the PFSS on a ring mentioned above. In the condensed phase, effectively non-local updates via the Metropolis algorithm accelerate the convergence to the stationary state, since it facilitates the melting of local, separated, metastable aggregates to finally merge to a single condensate.

Let us now discuss some examples.
Choosing $g_{ab}(m,n)\equiv p(m)$, i.e., making the two-point weight independent of the second argument and being the same for all links, one obtains $u_i(m_i|\dots)=u(m_i)\equiv p(m_i-1)/p(m_i)$. Here $u(m)$ depends only on the occupation number at the departure node.
If one now chooses $W_{ij}$ in any way that satisfies (\ref{kwij}), the stationary state
$w(\vec{m})= p(m_1)\cdots p(m_N)$ is precisely the steady state of a ZRP on an arbitrary network \cite{2006}.

A more complicated example is a one-dimensional closed chain so that $k_i=2$ for all sites $i$, and $g_{ab}(m,n)\equiv g(m,n)$ being independent of the link, but
depending on occupation numbers at its both ends. Equation (\ref{eq4}) then leads to
\bq
	u_i(m_i|m_{i-1},m_{i+1}) = \beta(m_i,m_{i+1}) \beta(m_i,m_{i-1}), \label{uipfss}
\eq
where $\beta(m,n)= g(m-1,n)/g(m,n) \label{beta}$. This is essentially the PFSS considered in \cite{evans1}. The only difference is that no assumption on the symmetry of $g(m,n)$ was made there. It has been shown in \cite{evans1} that when
\bq
	g(m,n)=\exp[-J|m-n|+U(\delta_{m,0}+\delta_{n,0})/2], \label{gevans}
\eq
in the thermodynamic limit the system exhibits a condensation transition above some critical density of particles $\rho=M/N$ which is a function of the parameters $J,U$. The nature of this condensate is different from an analogous phenomenon in the ZRP, because it emerges due to nearest-neighbor interactions and not due to the on-site potential $p(m)$ as in the former case. As a result, the condensate is extended over $\sim\sqrt{N}$ sites, see \cite{evans1} and \cite{bw1} for details.

Another simple special case is a square lattice with periodic boundary conditions on which we can specify hopping rates that factorize over adjacent sites.
Let us denote a site on this  two-dimensional torus by its coordinates $(i,j)$. From (\ref{kwij}) we obtain that
\ba
	W(i-1,j\to i,j)+W(i+1,j\to i,j) \nonumber \\ +W(i,j+1\to i,j)+W(i,j-1\to i,j) = 1,
	\label{w2d1}
\ea
so that the probabilities for ending up at target site $(i,j)$ should add up to $1$. Similarly, the probabilities for leaving a site $(i,j)$ as departure site should add up to $1$, that is
\ba
	W(i,j\to i+1,j)+W(i,j\to i-1,j) \nonumber \\ +W(i,j\to i,j+1)+W(i,j\to i,j-1) = 1.
	\label{w2d2}
\ea
Equations (\ref{w2d1}) and (\ref{w2d2})
are the only conditions on transition probabilities.
They have many solutions, which differ by the degree of current anisotropy, e.g., the current may flow only in horizontal, vertical, or in both directions.

\begin{figure}
	\includegraphics*[width=14cm,bb=0 0 1080 635]{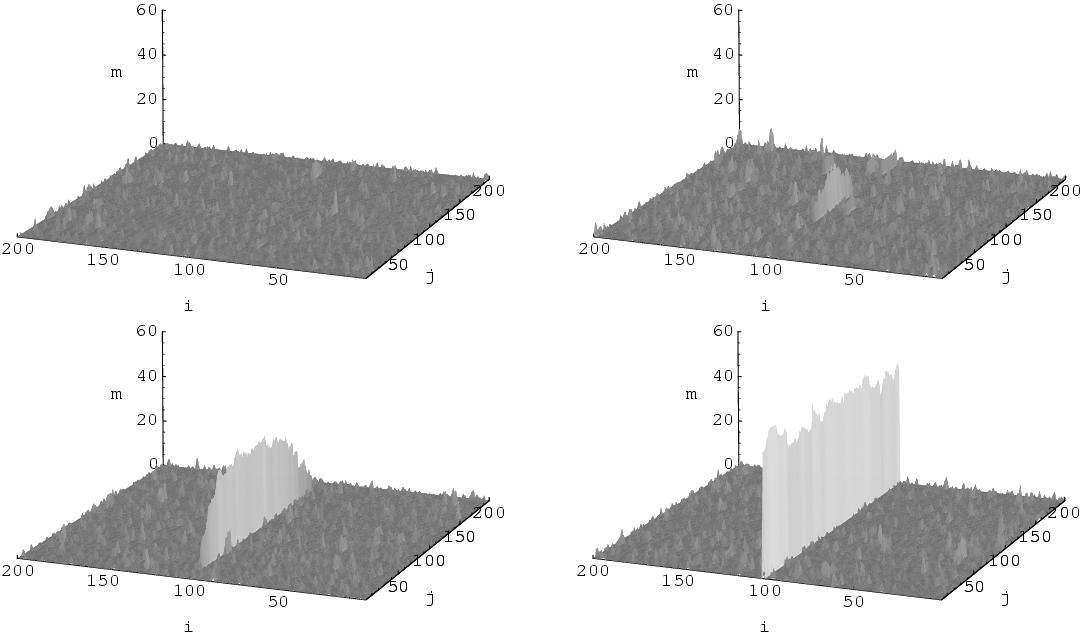}
	\caption{\label{fig1}Snapshots of Monte Carlo simulations of the anisotropic system with $L=200$, $J=U=1$, for various densities $\rho=0.2$ (upper-left), $0.3$ (upper-right), $0.35$ (bottom-left) and $0.5$ (bottom-right). For $\rho=0.2$ the system is in the liquid phase. For $\rho=0.3$ it is slightly above $\rho_1\approx 0.24$ --- a condensate emerges. For $\rho=0.35$ the condensate is fully developed. For $\rho=0.5>\rho_2\approx 0.45$, the borders of the condensate merge and the whole vertical line is uniformly covered.}
\end{figure}

\noindent{\bf Anisotropic interactions in two dimensions.}
Now we shall describe a case which (from a theoretical point of view) is of particular interest due to its analytic solvability and the existence of two phase transitions.
Let us assume that we have a two-dimensional lattice with $N=L\times L$ sites, periodic boundary conditions and the following steady state:
\bq
	w(\vec{m}) = \prod_{i=1}^L \prod_{j=1}^L g(m_{i,j},m_{i,j+1})\;\delta\left(\sum_{i,j}m_{i,j}-M\right).
\eq
 This implies that $g(m,n)$ lives on links only in one, say vertical, direction. On horizontal links, the weight is assumed for simplicity to be constant and equal to $1$, but in general one may choose any weight $f(m_{i,j})$ depending only on a single site. The hopping rate assumes now the form
\bq
	u(m_{i,j}|\dots) 
	= \frac{g(m_{i,j}-1,m_{i,j-1})}{g(m_{i,j},m_{i,j-1})}
	\frac{g(m_{i,j}-1,m_{i,j+1})}{g(m_{i,j},m_{i,j+1})},	
\eq
which is very similar to (\ref{uipfss}), but the particles can now jump either in horizontal or vertical direction. The current is defined by any $W$ obeying equations (\ref{w2d1}) and (\ref{w2d2}). Although we will not address dynamical issues here, for definiteness let us assume that the particles can jump only to the right with probability $W(i,j\to i,j+1)=p$ or to the top with probability $W(i,j\to i+1,j)=1-p$, while $W(i,j\to i,j-1)=W(i,j\to i-1,j)=0$. Moreover, let the weight $g(m,n)$ be chosen according to (\ref{gevans}).
Based on our knowledge on the separate ZRP and PFSS processes in one dimension, we would naively expect the following scenarios to be true: a) below a certain mass density, particles distribute over the two-dimensional grid homogeneously like a liquid; b) above some critical density, as a remnant of the interaction-dependent hopping in vertical direction, an extended condensate forms along each vertical line; c) above some critical density, as a remnant of the ZRP in one dimension, the overall condensate gets localized along a single vertical line, along which the form of the condensate is determined by the one-dimensional PFSS; d) the condensate stays localized along the horizontal axis, but looks like a liquid along the condensate-carrying vertical line.
Figure \ref{fig1} shows snapshots of Monte Carlo simulations for $J=U=1$ and increasing density $\rho$.
For our choice of weight functions we see that scenarios (a), (c) are realized below and above some $\rho_1$, respectively. Scenario (d) happens for some $\rho_2>\rho_1$, while scenario (b) is excluded. We shall derive next why only these scenarios can happen. We will argue that $\rho_1$ is the critical density of the corresponding PFSS in one dimension and also calculate $\rho_2$. Note that the whole system does not simply split into a direct product of a ZRP in one dimension and a PFSS in the other dimension, since both subsystems are coupled via the overall mass conservation.

First we observe the following. In the steady state we can treat the probability $w(\vec{m})$ as the weight of a microstate of a system being in equilibrium, therefore we can formally write the partition function of the system in the canonical formulation as
\ba\label{hi1}
	Z_{c2d}(N=L^2,M)&=&\sum_{\{m_{i,j}\}} \prod_{i=1}^L\prod_{j=1}^L g(m_{i,j},m_{i,j+1}) \delta\left(\sum_{i,j}m_{ij}-M\right)\nonumber\\
&=& \sum_{M_1=0}^M \cdots \sum_{M_L=0}^M \prod_{i=1}^L Z_{c1d}(L,M_i) \delta\left(M-\sum_{i}M_{i}\right),
\ea
where
\bq
	Z_{c1d}(L,\tilde{M}) = \sum_{m_1=0}^{\tilde{M}} \cdots \sum_{m_L=0}^{\tilde{M}} \prod_{i=1}^L g(m_i,m_{i+1}) \delta\left(\tilde{M}-\sum_i m_i\right) \label{eq:z1d}
\eq
is the partition function of a 1d PFSS. Since (\ref{hi1}) has the same functional form as the partition function for the zero-range process \cite{evans0501338}, the partition function $Z_{c1d}$ may be seen as the weight $p(m)\equiv Z_{c1d}(L,m)$ that is now associated with the total mass $m=M_i$ along the $i$th vertical line.

In \cite{bw1} we have shown that for $m>\rho_c L$, where $\rho_c$ is the critical density for condensation in the one-dimensional system, $Z_{c1d}(L,m)$ behaves as $\sim \exp(-c\sqrt{m})$ with some $c>0$. This means that for large $m$, the hopping rate of the corresponding ZRP, $u(m)=p(m-1)/p(m)=Z_{c1d}(L,m-1)/Z_{c1d}(L,m)$, behaves as $u(m)\cong 1+c/(2\sqrt{m})$. For such a hopping rate it is known \cite{evans0501338} that the ZRP exhibits a condensation transition. The condensate occupies a single site and the fluid-phase distribution is a stretched exponential distribution.
Since $m$ denotes now the mass along a vertical line, the mapping from the ZRP back to the anisotropic model allows us to predict the spontaneous symmetry breaking into the state with a condensate so that one of the masses $M_i$ will grow to pick up all the difference $\Delta M = M-\rho_1 L^2 = L^2(\rho-\rho_1)$ between the total mass $M$ and the mass $\rho_1L^2$ in the critical background.
So, differently from scenario (b), not each line carries its own condensate, but the condensate is localized onto a single line if there is a condensate at all, while all other lines have a mass density close to the critical value.
To predict the critical density $\rho_1$ we observe that (\ref{hi1}) in the grand-canonical formulation,
\ba\label{hi2}
	Z_{g2d}(L,z) &=& \sum_M Z_{c2d}(L,M) z^M = \sum_{\{M_i\}} \prod_{i=1}^L Z_{c1d}(L,M_i) z^{M_i} \nonumber \\
	&=& \left(\sum_M Z_{c1d}(L,M) z^{M}\right)^L,
\ea
becomes just a power of a grand-canonical partition function in one dimension. Since the condensation transition is determined by the non-analytic behavior of $Z_{g2d}$, the critical density $\rho_1$ must be the same as $\rho_c$ of the one-dimensional PFSS. For $J=U=1$ it reads $\rho_1\approx 0.24$, as follows from \cite{bw1}.

In order to check how the particles are distributed along the line which carries the condensate, in particular to determine the shape and extension of the condensate, we will again make use of results of \cite{bw1} for the one-dimensional system. It is proved there that for sufficiently large systems, the condensate has a quasi-parabolic shape with very sharp borders. Its width $W$ grows proportionally to $\sqrt{\Delta M}$, so that here $W=w_0 L\sqrt{\rho-\rho_1}$ with some constant $w_0$ that depends only on $J,U$ from (\ref{gevans}) and reads $w_0\approx 2.2$ for $J=U=1$. This is precisely the shape seen in figure \ref{fig1}, for $\rho_1<\rho<\rho_2$. When the density exceeds $\rho_{2}=1/w_0^2 +\rho_1\approx 0.45$, the width $W$ becomes equal to the linear size $L$. Since fluctuations of occupation numbers in the condensate can be neglected in the thermodynamic limit \cite{bw1}, for large systems and $\rho>\rho_{2}$ there are no empty sites. The ultralocal weight $e^{U(\delta_{m,0}+\delta_{n,0})/2}$ becomes equal to one and hence $g(m,n)$ effectively behaves as $\exp(-J|m-n|)$ as if $U$ were zero.
We may therefore see the line carrying the condensate as a new PFSS with $U=0$. From \cite{evans1} we know that such a system is always in the liquid state. This means that
at $\rho=\rho_{2}$ the system undergoes a second (geometric) phase transition to a state, in which both borders of the condensate merge and the particle distribution along the condensate line looks uniform apart from fluctuations, see figure \ref{fig1}, which corresponds to scenario (d). We call it a geometric phase transition as the condensate percolates all over the line. It is not just a finite-size effect of merging borders, because $\rho_2$ remains finite also in the thermodynamic limit $L\to\infty$.
Thus the condensate changes its shape from quasi-parabolic in vertical and localized in horizontal direction, to a homogeneous distribution vertically, but still localized horizontally. Any transient remnant of the former condensate peak gets rapidly washed out towards a homogeneous distribution.

In summary, we proposed a class of hopping rates that leads to pair-factorized steady states on arbitrary graphs. The proof holds in particular for non-local and inhomogeneous hopping rates. As an example we studied an anisotropic two-dimensional system with PFSS, for which we predicted the onset of condensation including the shape and the scaling of the condensate by dimensional reduction of the
two-dimensional system to an effective zero-range process in one dimension.

\ack{B.W. and W.J. thank the EC-RTN Network "ENRAGE" under grant No.~MRTN-CT-2004-005616 for support.
B.W. would like to thank the International Center for Transdisciplinary Studies (ICTS) at Jacobs University for its hospitality and support of several visits during this collaboration.
}

\end{document}